# Harnessing AI for Speech Reconstruction using Multi-view Silent Video Feed

Yaman Kumar
Adobe Systems
Noida
ykumar@adobe.com

Mayank Aggarwal
Netaji Subhas Institute of Technology
New Delhi
mayanka.mp@nsit.net.in

Pratham Nawal
Netaji Subhas Institute of Technology
New Delhi
prathamn.mp@nsit.net.in

Shin'ichi Satoh
National Institute of Informatics
Tokyo
satoh@nii.ac.jp

Rajiv Ratn Shah
Indraprastha Institute of Information Technology Delhi
New Delhi
rajivratn@iiitd.ac.in

Roger Zimmermann
National University of Singapore
Singapore
rogerz@comp.nus.edu.sg

## ABSTRACT

Speechreading or lipreading is the technique of understanding and getting phonetic features from a speaker's visual features such as movement of lips, face, teeth and tongue. It has a wide range of multimedia applications such as in surveillance, Internet telephony, and as an aid to a person with hearing impairments. However, most of the work in speechreading has been limited to text generation from silent videos. Recently, research has started venturing into generating (audio) speech from silent video sequences but there have been no developments thus far in dealing with divergent views and poses of a speaker. Thus although, we have multiple camera feeds for the speech of a user, but we have failed in using these multiple video feeds for dealing with the different poses. To this end, this paper presents the **world's first ever** multi-view speech reading and reconstruction system. This work encompasses the boundaries of multimedia research by putting forth a model which leverages silent video feeds from multiple cameras recording the same subject to generate intelligent speech for a speaker. Initial results confirm the usefulness of exploiting multiple camera views in building an efficient speech reading and reconstruction system. It further shows the optimal placement of cameras which would lead to the maximum intelligibility of speech. Next, it lays out various innovative applications for the proposed system focusing on its potential prodigious impact in not just security arena but in many other multimedia analytics problems.

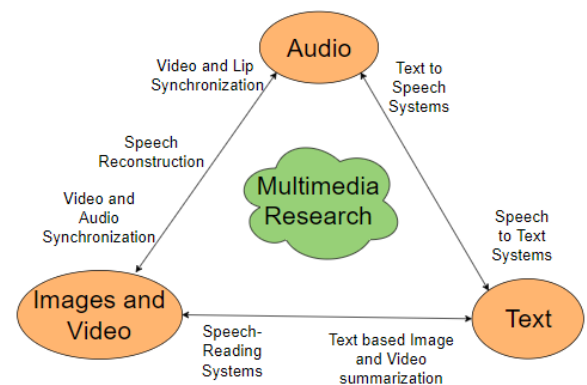

**Figure 1: Domain conjugation and media interaction in multimedia research**

## 1 INTRODUCTION

Multimedia as a research topic is more than just a combination of diverse forms of data [53], *i.e.*, audio, video, text, graphics, *etc*. Essentially, it is the amalgamation and the interaction among these divergent media types that lead to the formation of some extremely challenging research opportunities such as that of speech reading and reconstruction which are presented in this work [54]. Figure 1 introduces the various domains in multimedia research. Traditionally, multimedia system have been grouped into text, audio and video based information systems. This paper revisits one of the most fundamental problems in multimodality, that of speech reading and reconstruction systems. This problem belongs to one of those research arenas which are cross domain and exploit the full breadth and depth of multimedia research since it involves not just speechreading but the synchronization of video with lip movements as well as reconstruction of the audio.

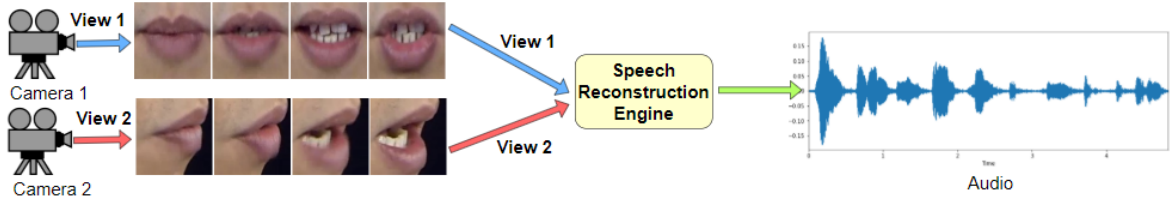

Figure 2: The model proposed for speech reconstruction with its inputs, output and the process

Speechreading involves looking, perceiving, and interpreting spoken symbols. Speech is considered to be composed of phonemes which are the smallest distinct and detectable units of sound in a given language [55]. Phonemes are produced by different movements of teeth, tongue, lips, *etc.* Visemes, on the other hand, are facial images including jaw and lip movements which are used to define and generate sound [10]. The task of speechreading is made difficult by the fact that often several phonemes correspond to a single viseme, thus producing ambiguity when trying to infer speech from visemes only. For example, it is very difficult to distinguish the characters *'p'* and *'b'* just by looking at the image of a speaker who speaks both the characters. That is because characters like *'p'* and *'b'* belong to the same viseme class. Similar is the case with expressions like *"Elephant Juice"* and *"I love you"* which though having similar visemic appearances definitely have very different sounds and meanings [26].

In the recent years, Automatic Speech Recognition (ASR) systems have gained significant traction, with systems being deployed widely in cars, mobile phones, homes, *etc.* [18, 62]. However, considering settings such as that of a car on a road, it is not ideal and is plagued by very low signal to noise ratio [56]. This leads to speech recognition systems failing utterly in the absence of usable audio[32]. However, these fallacies with ASR systems can be solved satisfactorily by deploying speech reading and reconstruction systems which can augment the understanding of ASR systems or can even reconstruct the speech for them. With the advent of multiple frontal cameras in embedded devices like mobile phones, webcams and cars, additional information in the form of visual output of speech such as lip movements is easily accessible. Currently, the visual feed thus obtained is being ignored or wasted. This can be effectively integrated with ASR system via the use of technology proposed in the current paper. Another crucial case is of venues requiring critical security infrastructure such as cockpits, battlefields and even public places such as railway stations, airports, *etc.* In this case, the cameras either do not record the audio or the audio recorded is too feeble or jagged to be of any use. This situation unequivocally calls for a solution. Additionally, considering security and crime scenarios themselves, due to the availability of camera footage, visual cues such as lip-movements have been used in the past for solving deleterious crimes [1–5, 15, 42]. However, professional lipreaders are required for rendering their services in these cases, which are not only expensive but highly limited. The single solution for all these challenges is a system having speech reading and reconstruction capabilities which can also effectively integrate with ASR systems.

Some of the earliest works reported in the field of lipreading are those by Potamianos et al. [47] and Lan et al. [30]. Recent works such as those by Cornu and Milner [17] and Akbari et al. [6] perform the task of lipreading on the GRID database [16]. Where as the previous works [6, 17] focused on single view based lipreading often with hand-picked features, this work leverages the power of multiple views for the task at hand. Thus these are the key contributions of the present work:

(1) The authors build the **world's first ever** intelligible speech reading and reconstruction system which uses a multi-view silent visual feed to generate non-robotic (humanistic) synchronized audio for silent videos. The overview of the system consisting of inputs, outputs and the process is given in Figure 2. As noted by Zhou et al. [64], pose forms a critical bottleneck for all the systems developed till now. In a practical situation, it cannot be assumed that users would continually face the camera while speaking. Normally, users twist and turn in front of cameras. Thus the various poses captured in the feed create a serious challenge for all practical ASR systems.
(2) They answer the pertinent question of the optimal placement of cameras with the problem of lipreading in sight. They show the positions of cameras which would help to produce a highly reliable audio from a silent or noisy video.
(3) The authors show a model for generating audio which is in sync with the lip movement of the speaker. However there have been many previous attempts at lipreading which generated text transcripts [31, 33, 46, 66] and one could argue, that from those texts, one can generate an audio. However the audio thus generated would not be in sync with the video, thus creating non-coherence and an unnecessary lag (or haste) between what is spoken and what is seen. This immediately makes it unsuitable for its use in real-time applications.
(4) In addition, another disadvantage of using the audio that could be generated by existing lipreading systems [31, 33, 46, 66] which first lipread thereby generating text, and then convert text to audio, is that they generate a robotic sound carrying no coherence with the actual speech of a speaker. This is a serious challenge for the existing systems considering the fact that the system proposed in this paper could also be used in multimedia applications such as video conferencing and human-computer interaction systems where the environment demands actual sound and speech of a speaker. Thus when the existing lipreading models would fail, the one proposed could be deployed successfully.

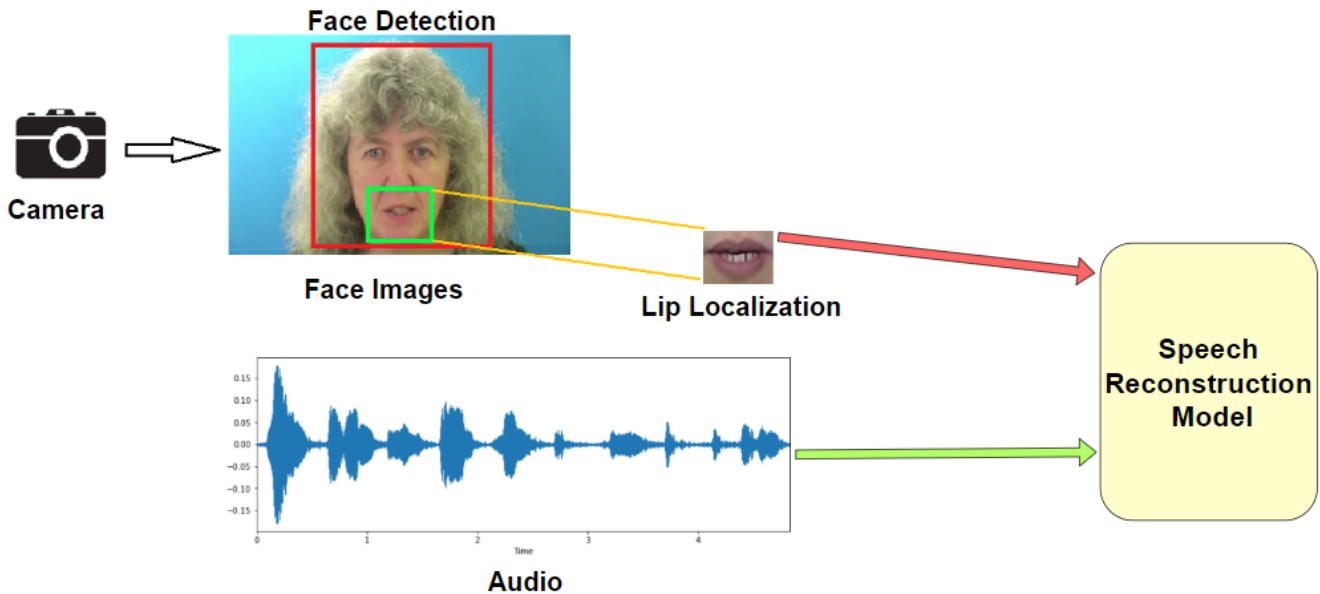

Figure 3: Training of the speech reading and reconstruction model. Input to the model is silent video feed and the actual audio of the speech

(5) Additionally and most importantly, the existing lipreading systems are text based systems, which are highly dependent on language and vocabulary. If language and vocabulary are changed, they fail utterly. A talk between bilingual speakers cannot be understood by the existing systems. Here the authors propose a system which is not only vocabulary but language independent too. They have not used any language specific feature in this implementation thus making the system independent of any and all languages.

(6) In light of the current work, the authors discuss several closely associated and allied tasks such as augmentation of audio in movies, helping people with and without hearing disabilities to understand speech. These applications can greatly benefit from building a speech reading and reconstruction system like the one proposed in Figure 2.

The rest of the paper is organized on these lines: Section 2 presents the previous work in this domain. Section 3 discusses the approach followed, overall methodology and models used by the authors in designing the system. Section 4 shows the results for the experiments carried out. In Section 5, the authors note the possible futuristic application areas which can greatly benefit from the current work. Finally, Section 6 concludes the paper while presenting the work intended to be undertaken in the future.

## 2 RELATED WORK

The documentation for lipreading spans centuries with some of the earliest works being reported in the 17th century [12]. It mentions that useful information can be extracted from facial movements of a speaker. Several psychological studies have demonstrated that lower level additional information helps in hearing [19, 20]. Experiments and research studies have shown that people with [11, 35] and without [59] hearing impairment use visual cues for understanding and augmenting the understanding of what a speaker is trying to say. As noted by Chen and Rao [13], skilled lip-readers perform the task by looking at the configuration and movement of tongue, lips and teeth.

One of the first audio-visual recognition systems was developed by Petajan [44] in which the camera was used to capture the mouth images which were later thresholded into black and white. Binary images thus obtained were analyzed to derive open mouth area, height, perimeter, *etc*, which were later used for recognition. Traditionally, lipreading has been considered as a classification task where either words or phrases are identified and selected from a limited lexicon. Some of the earliest works using this were those by Ngiam et al. [37], Noda et al. [38], Petridis and Pantic [45]. Authors in those works used hand-picked features used with a combination of deep learning models followed by classifiers. Several end-to-end deep learning models have also been developed which rely on a combination of convolutional and recurrent neural networks [8, 14, 61].

Many have previously ventured into taking the tasks of lipreading for digits only, with some of the notable works being Chen and Rao [13], Pachoud et al. [41], Sui et al. [58]. However these systems, since being very specific face a lack of generality.

In their work, Cornu and Milner [17] used hand-engineered features to reconstruct audio from video, using a deep-learning network. This method then was modified by Ephrat et al. [21] who used CNN over the entire face of the speaker. Both of these works used Linear Predictive Coding (LPC) [22] coefficients for modeling audio features.

As noted in Section 1, it is impractical to assume that speakers would be still and continuously facing a camera in a singular pose for their full speech, which is the common unsaid, unwritten assumption of almost all the previous works mentioned. This seriously impedes their lipreading model's ability when the subject turns to a new pose. Thus camera view and pose can drastically affect the model's accuracy in predicting and deciphering the speech. Thus with regards to taking pose into consideration, there have been three common approaches thus far:

(1) Many authors have used non-frontal views to extract pose dependent features and then used them for building ASR. For instance, Iwano et al. [25] used profile views and then calculated optical flow for each image. In their work, Lucey and Potamianos [33] showed that using a profile view was inferior to using a frontal view when results were derived using their ASR pipeline. However, Kumar et al. [28] used their geometric pose dependent features and other hand-extracted features to recognize the isolated words, hence showing that results from profile views were better than frontal views.
(2) Some research projects have used pose dependent features and transformed them into pose independent features, then eventually using them for classification. For instance, Lucey and Potamianos [33] and Lucey et al. [34] transformed pose dependent features from an unwanted camera view to the wanted one. However, the results showed that the performance was substantially worse when compared to the same pose views. In the work by Lan et al. [29], it was inferred from the experiments on the derived pose dependent features that 30° (degrees) (from the frontal view) was the optimal view for the task of lipreading. Then the features were projected to the optimal view which was then tested on other views. Extracting features from pose dependent features, though it seems to be a viable approach, suffers from a grave lack of data for the large number of possible views. Pose independent features based systems being view-blind, suffer from a huge drop of accuracy and performance [64].
(3) To the best of the authors' knowledge, there have been only four previous works who have attempted to build multi-view lipreading systems generating **text** from silent/noisy video feeds. Lucey and Potamianos [33] used discrete cosine transform based features subsequently fed to a Hidden Markov Model as input to generate text for the speech. They showed that a two view system outperformed a single view system. Lee et al. [31] used various CNN-LSTM based architectures to generate text from speech. They tested the combination of all the five views which were subsequently found out to be worse than all the single views. Zimmermann et al. [66] used Principal Component Analysis along with a LSTM and HMM based architecture on OuluVS2 to obtain speech transcripts. They too showed that combining all views led to worse performance. Finally, Petridis et al. [46] used CNN-LSTM based networks to predict text from video sequences on OuluVS2 dataset. The accuracy they obtained was close to 96% using multiple views. It is worth noting that none of the said works have ventured into the task of producing audio from a silent video and as noted in Section 1, thus seriously curtailing the deployment of these systems in real time multimedia based applications which require sync with videos and the original human voice. Additionally, usage of these systems creates language and vocabulary training dependencies on the model.

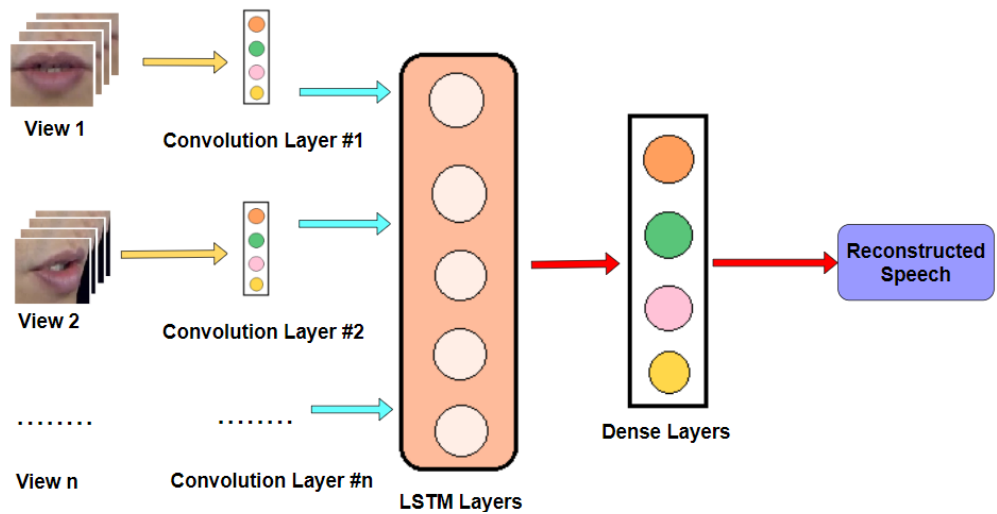

Figure 4: CNN-LSTM based architecture used for speech reading and reconstruction

## 3 METHODOLOGY

The literature survey suggests that researchers have spent great efforts and time to find the visual features associated with lips, mouth and teeth to map facial features to speech. The authors in this paper bypass that requirement by employing a Convolutional Neural Network (CNN) based deep learning model. A high level view of the system is shown in Figure 4. It takes feeds from multiple cameras and then produces the audio corresponding to the images obtained.

Given a speaker's images, a CNN-LSTM based network automatically learns the optimal features required for reconstructing the audio signal.

This model has the advantage of not depending on a particular segmentation of the input feed into words or letters or even phrases and sentences. It uses "natural supervision" to learn a prediction target from a natural signal of the world [40]. A large enough dataset can train this network successfully to predict out-of-vocabulary words (OOVs).

### 3.1 Audio Features

Many techniques such as sinusoidal models, vocoders, *etc*, have been employed for speech reconstruction [36, 57]. One could also work directly with spectrograms but as per the experiments carried out by the authors, the quality of the audio reconstruction obtained from it is poor. Results obtained from those experiments are not presented in this work due to their inferior quality. Raw audio waves cannot be used due to a lack of suitable loss function. The authors in this work have chosen Linear Predictive Coding (LPC) for representing the audio speech [22]. LPC is a technique to represent the compressed spectral envelope of speech using a linear predictive model. It produces high-quality speech using low bit-rate. The order P of LPC was varied and chosen so as to get the best quality speech. Line Spectrum Pairs (LSPs) [24] can represent LPC coefficients in a robustly quantized manner. Thus LSPs prove to be useful at representing LCP for the task at hand. The current audio pipeline has the advantage of modeling other parts of human speech such as emotion, prosody, *etc.*

### 3.2 Training the Model

The proposed system is shown in the Figure 4. It consists of a CNN-LSTM based architecture. While CNN layers extract the visual features from images, LSTM layers are used for taking into consideration the time dependence of speech (both video and audio). Finally dense layers are then connected to LSTM layers which produce audio corresponding to the sequence of images.

As shown in the Figure 3, for training of the model, the system requires multiple views of the same subject speaking something and the corresponding audio. The images are preprocessed in order to obtain better features from the images. The preprocessing involved converting the images to gray-scale, resizing them and then applying Contrast Limited Adaptive Histogram Equalization (CLAHE) [51]. The obtained images are then fed to the CNN-LSTM based network.

The output of the CNN-LSTM architecture is the audio corresponding to the video given. It is processed according to the technique mentioned in Section 3.1.

Although the goal is to map a single video frame $I_i$ to a single audio vector $S_i$, due to instantaneous lip and mouth movements, it would be inefficient for the system to do so. Thus the system maps a sequence of images to a part of the corresponding audio vector. It serves as a temporal neighborhood for the $I_i$ image and has a corresponding $S_i$ audio vector. The sequence length of images was taken as a hyper-parameter and was optimized to get the best quality sound. Thus the timesteps parameter of LSTM was obtained as 5 images. In other words, the model considered 5 previous frames to be the temporal dependency of a frame while constructing the speech. The network was then trained on a combination of mean squared and correlation based error loss metric.

Table 1: Results for the proposed model. V1, V2, V3, V4, V5 represent the single views mapping to 0°, 30°, 45°, 60° and 90° respectively. V1_2 means the combination of 0° and 30° and V1_4 represents the combination of 0° and 60°. Results shown are PESQ values for two speakers- Speaker 1 (Male) and Speaker 10 (Female)

| View | Speaker 1 | Speaker 10 |
|---|---|---|
| **V1** | 1.7674 | 1.7812 |
| **V2** | 1.7255 | 1.7129 |
| **V3** | 1.6161 | 1.6012 |
| **V4** | 1.4962 | 1.5312 |
| **V5** | 1.4284 | 1.4453 |
| **V1_2** | 1.9683 | 1.9978 |
| **V1_4** | 1.8636 | 1.9463 |

## 4 EVALUATION

### 4.1 Database Used

In the current work, the authors use the OuluVS2 database [7]. OuluVS2 is a multi-view audio-visual database. It consists of 53 speakers with five different views spanning between frontal and profile views. The talking speed was not regulated and depended on the speaker. The database consists of three different types of data:

(1) Ten short English phrases such as "Good-Bye" and "Hello". These phrases were also present in OuluVS [63] database which was extensively used in various studies [39, 43, 49, 50, 65].
(2) Ten random digit sequences uttered by 53 speakers.
(3) Ten randomly chosen TIMIT sentences. [67]

A variety of speaker appearance types were considered in the dataset with most of them being of the following types: European, Chinese, Arabian, African and Indian/Pakistani. Cameras recorded these subjects from 5 different angles: 0°, 30°, 45°, 60° and 90°. Thus this database serves well for this work. The database consists of speech which ranges from digits to commonly used phrases and TIMIT sentences [67]. The speakers have different ethnic appearances and importantly, were recorded from 5 different views, thus forming the base for our subsequent analysis.

### 4.2 Evaluation Metric used

In this work, Perceptual Evaluation of Speech Quality (PESQ) [52] is taken as the metric used to evaluate the quality and intelligibility of the speech. While there are many choices for evaluation of speech quality, the authors have chosen this since it corresponds well to the perceived audio quality [23], in addition to the system's loss function. Among all the measures such as PSQM [9] and PEAQ [60], PESQ has been recommended by ITU-T for speech quality assessment of 3.2 kHz codecs [48]. PESQ compares two audios one of them being original and the other one being generated by the system. The PESQ system first level aligns the two audios, then after passing them through filters, time aligns them, passes them through an auditory transform and finally extracts two distortion parameters. These distortion parameters are the difference between the transform of the two signals. Finally after aggregating the signals in frequency and time, they are mapped to a mean opinion score. A PESQ score ranges from -0.5 to 4.5 with 4.5 being the score for high quality reconstruction of the original audio. Previous works such as Ephrat et al. [21] and Akbari et al. [6] also use this measure.

Table 2: Average PESQ results after comparing original audio with the audio obtained as explained in Section 3.1. Results are for the two speakers- Speaker 1 (Male) and Speaker 10 (Female)

| Speakers | PESQ |
|---|---|
| **Speaker 1** | 2.5291 |
| **Speaker 10** | 2.6255 |

The authors would like to note that using the audio pipeline as explained in Section 3.1, the audio constructed was compared to the original audio and the results obtained are shown in Table 2.

### 4.3 Results

Table 1 contains the results obtained using the speech reconstruction model used. It shows both the single view and multi-view results for speaker 1 (a male) and speaker 10 (a female) of OuluVS2. The speakers 1 and 10 have been chosen randomly, and not because of some specific strategy. One of them, *i.e.*, speaker 1 is a male and the other one, *i.e.*, speaker 10 is a female, thus demonstrating the results for both the sexes. Due to the paucity of space, only the best results among multiple views are shown in the table. View 1 corresponds $0°$, view 2 to $30°$, view 3 to $45°$, view 4 to $60°$ and view 5 to $90°$. As shown by the table, views 1 and 4 and views 1 and 2 outperform other multiple as well as single views.

For the male speaker, the combination of views 1 and 2 shows an improvement of approximately 26% over view 1 and 33% over view 2 individually while the combination of views 1 and 4 shows an improvement of 12.5% over view 1 and 80% over view 4. For the female speaker, the combination of views 1 and 2 shows an improvement of approximately 27% over view 1 and 40% over view 2 individually while the combination of views 1 and 4 shows an improvement of 21% over view 1 and 80% over view 4. This is a very significant improvement considering the fact that the cost of including already available information (in terms of multiple video feeds) into the model is negligible where as the gains are outstanding. These preliminary results show that the model performs equally well for male as well as female speakers.

Thus considering the task of placement of cameras, the model would perform best if the angle between the two cameras is in the range of $30°$ to $60°$. This would give the maximum intelligibility of speech and can help immensely in the reconstruction of speech.

## 5 DISCUSSION AND APPLICATIONS OF LIPREADING SYSTEMS

As illustrated in Figure 5, there are a host of applications for the current multi-view lipreading system:

- Lipreading in general and multi view lipreading in particular can be used to augment the understanding of what a person says. It is highly useful for hearing impaired [11, 35] people. Thus a major use case of the multi-view lipreading system would be hearing impaired people who would largely benefit with accurate text of what the speaker is speaking.
- People who have full hearing capabilities [59] can employ this system in their day-to-day lives. For instance, as noted by [13], listeners face trouble comprehending the audio speech when either the speaker has a foreign accent or even when the speaker is unfamiliar to the listener. This problem can be solved using multi-view lipreading. Ultimately, the audio signals of foreign or accented speakers might be different but the lip-movements required for producing the standard alphabet are the same. Thus while the video sequence would be from the foreign speaker and would remain the same but the audio could be changed. The audio used in the training of the system can be replaced by an audio which is familiar to the listener, thus while the words would be of the foreign speaker the sound would be of a familiar person thus making idea exchange much more efficient.
- One of the major deployment opportunities of the multi-view lipreading are crowded places whose security is of grave concern to the public at large. Huge amounts of money, manpower and systems are rightly deployed in recording, listening, surveilling, patrolling and in general protection and security related activities. At crowded public places like railway stations, airports, road intersections, stadiums, auditoriums, *etc.* it becomes imperative for the authorities to look for and keep a watch on unwanted elements. For this, CCTV and camera footage become a vital source of information of their movements. The speech-reconstruction system proposed in this paper, if deployed, with multiple camera feeds can become an indispensable tool in the hands of authorities.
- The other major use case for the system proposed is movies. Movies, whether in the format of entertainment or for educational purposes such as in MOOCs (Massively Open Online Courses) can reap the benefits of this system. Often multiple cameras record the subject in a movie simultaneously out of which a single view is then shown to the viewers. Whenever the speech of a subject is non-audible or non-decipherable, the viewers are left wondering what the speaker said. At this point of time, such an approach is not only highly inefficient but wasteful too. The feed from multiple cameras can easily be used to augment the speech and the sound of the speaker.
- Highly noisy environments such as those of parties or festival celebrations serve as another area of deployment of the system proposed. Multiple cameras recording the subject can inform the listener of the speech in real time.
- With the advent of mobile devices and multiple front cameras that come along with them, applications like video conferencing have seen a tremendous rise in their user-base. The authors think of these as a great platform for the employment of the system proposed due to the near perfect camera feed, most of which contains only the speaker's face. Due to bad reception, it is often the case that audio becomes jagged. Thus the speech reconstruction system proposed

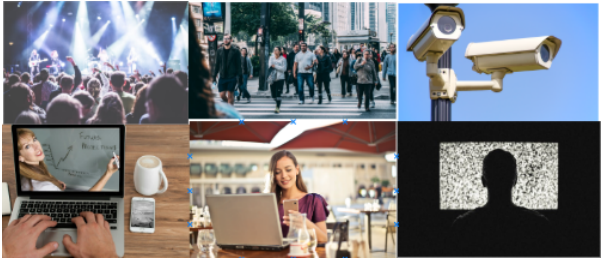

Figure 5: Applications of multi-view lipreading. The applications shown are (starting from the top-left in clockwise order) in noisy environments such as parties, at public places such as road intersections, for enhancing security infrastructure such as by making use of multiple camera at airports, in augmenting audio of MOOCs, in correcting audio in video conferencing applications and in movies

- can reinforce the speech of the speaker using the video feed available.
- One of the very worthwhile applications of the system proposed is in the medical field. It can be a boon for people with cerebral palsy. Cerebral palsy is a condition which makes it hard to move certain parts of the body. The pathology also affects the communication skills of the patient. In the US alone, it affects 764,000 people [27]. The system proposed can greatly affect the lives of people suffering with cerebral palsy in a very positive way. The system deployed on a camera can detect their lip movements and then construct audio for those movements. The system, for this application would be trained on audio taken from a healthy speaker and video sequences from the patient. The lip sequences thus learned from the patient would be mapped onto the audio of a healthy person, thus giving him a new voice. This would immensely help the patient in communicating effectively with the world.
- A very interesting application of this system could be silent videos (both movies and documentaries) recorded in $19^{th}$ and $20^{th}$ centuries. Lipreading followed by speech reconstruction performed on these videos could potentially introduce audio in these movies. This could revive a new wave of interest in these movies.
- A highly contemporary application of this system could be various gaming systems which depend on human computer interaction. Human interaction in the form of speech can be augmented and enhanced using this system.

## 6 CONCLUSIONS AND FUTURE WORK

The paper presented a model for generating intelligent speech leveraging silent feeds from multiple cameras at different angles. It also showed the optimal placement of cameras which would lead to the maximum intelligibility of the speech. To the best of the authors' knowledge, this is the *first ever* model built for doing so. The authors' finally laid out some potential industrial and academic applications which could benefit tremendously from the proposed system. The model was evaluated on a multi-view multi-speaker audio-visual database, OuluVS2, for its testing and training purposes.

The short term future goal for this work is to use more views for the task of speech reading and reconstruction. In the present work, the authors presented preliminary results which proved the concept that speech reconstruction could be made better if one can leverage the power of multiple views. The authors in this work used just two views for this purpose, but they intend to carry forward the task of speech reading and reconstruction for more number of views. Additionally, they also look forward to extend the model for words which are out of the limited lexicon of OuluVS2. This is a crucial task in light of the various potential applications presented in Section 5.

The long term aims and objectives in this research arena are immense. Before being able to actually commercialize speech reading and reconstruction systems, the research has to improve these systems drastically. Currently all the research is conducted on speakers in laboratory conditions with ample lighting facilities. In the real world, this is not always the case and the lighting is often poor and diffused. Practical lipreading and speech reconstruction systems have to overcome these hitches. They need to be useful in not just room and lab environments but at real places like airports and railway stations. In addition, for these models to be deployed on camera chips, they need to be smaller in size. Currently not only are the systems' size huge but the processing power required to train and test is colossal when compared with the processing power of camera chips. This is a serious hindrance for the deployment of these immensely helpful systems at a large scale.

## ACKNOWLEDGMENTS

This research was supported in part by the National Natural Science Foundation of China under Grant no. 61472266 and by the National University of Singapore (Suzhou) Research Institute, 377 Lin Quan Street, Suzhou Industrial Park, Jiang Su, People's Republic of China, 215123.

We gratefully acknowledge the support of NVIDIA Corporation with the donation of a Titan Xp GPU used for this research